\documentclass[aps,preprint,superscriptaddress,nofootinbib]{revtex4}

\usepackage{amsmath}
\usepackage{graphicx}
\usepackage{color}
\usepackage{slashed}

\begin{document}
	
\title{Cosmic Censorship and Weak Gravity Conjecture in the Einstein-Maxwell-dilaton theory}

\author{Ten-Yeh Yu}\thanks{%
	E-mail: coteric805@icloud.com}
\affiliation{Department of Physics and Center for High Energy Physics, Chung Yuan Christian University, Chung Li City, Taiwan}

\author{Wen-Yu Wen}\thanks{%
	E-mail: wenw@cycu.edu.tw}
\affiliation{Department of Physics and Center for High Energy Physics, Chung Yuan Christian University, Chung Li City, Taiwan}
\affiliation{Leung Center for Cosmology and Particle Astrophysics\\
	National Taiwan University, Taipei 106, Taiwan}

\begin{abstract}
We explore the cosmic censorship in the Einstein-Maxwell-dilaton theory following Wald's thought experiment to destroy a black hole by throwing in a test particle.  We discover that at probe limit the extremal charged dilaton black hole could be destroyed by a test particle with specific energy.  Nevertheless the censorship is well protected if backreaction or self-force is included.  At the end, we discuss an interesting connection between Hoop Conjecture and Weak Gravity Conjecture.
\end{abstract}

\maketitle


	
\section{Introduction}

The Weak Cosmic Censorship Conjecture asserts that singularities formed by gravitational collapse of matter are hidden behind event horizons \cite{Penrose:1969pc, Penrose:1964wq}.  Its basic idea is to resolve the apparent conflict between generic occurrence of singularity and smooth geometry in classical general relativity.  In a seminal work, Wald proposed an illuminating gedanken experiment to test this conjecture by plunging a test particle into an extremal Kerr-Newman black hole \cite{Wald:1974wa,Wald:1997wa}. Wald first showed that in order to expose the singularity, the particle must carry high charge and angular momentum to mass ratio.  On the other hand, he showed that the very particle cannot be captured by an extremal black hole, therefore the censorship is protected.  Wald's work has received lots of attention and followed by extensive studies \cite{Hubeny:1998ga,Hod:2002pm,Matsas:2007bj,Jacobson:2009kt, Jacobson:2010iu,Chirco:2010rq,Gwak:2011rp, Saa:2011wq,Gao:2012ca,Li:2013sea,Duztas:2013wua,Duztas:2013gza,Siahaan:2015ljs,Semiz:2015pna,Toth:2015cda,Gwak:2015fsa,Gwak:2015sua,Cardoso:2015xtj,Revelar:2017sem,Sorce:2017dst,Duztas:2017lxk,An:2017phb}. In some cases, such as nearly extremal RN \cite{Hubeny:1998ga,Matsas:2007bj} or Kerr black holes \cite{Jacobson:2009kt, Jacobson:2010iu, Saa:2011wq}, a test particle were able to destroy the horizons and censorship could be violated.  Although it is widely believed that the censorship could be restored if self-energy(force) correction \cite{Hod:2002pm}, backreaction or radiation effect were considered properly, there is no conclusive proof.  Instead, it is a nontrivial question to ask that in which background or to what extend the weak cosmic censorship conjecture is applicable.

This article concerns the cosmic censorship in the Einstein-Maxwell(-axion)-dilaton theory, which serves as a simple model of compactified low-energy string theory\cite{Duff:1994jr,Duff:1996qp}.  Specifically, we shall consider a consistently-truncated four-dimensional model obtained by toroidal compactification of the low-energy heterotic string theory, whose action reads ($G=c=1$)

\begin{equation}\label{eqn:action}
{\cal L}=\frac{\sqrt{-g}}{16\pi} \big[ R-2(\nabla \phi)^2 -e^{-2a\phi}F^2 \big]
\end{equation}

This model describes a massless scalar field in coupled with the abelian vector field.  In this paper, we only consider the dilaton black hole\cite{Gibbons:1982ih,Garfinkle:1990qj} with a specific choice of scalar-Maxwell coupling\footnote{The $a=0$ case yields the Reissner-Nordstr\"{o}m solution, while $a=\sqrt{3}$ case corresponds to the Kaluza-Klein black hole.} $a=1$, known as the Gibbons-Maeda-Garfinke-Horowitz-Strominger (GMGHS) solution.  The dilaton plays nontrivial role to replace the would-be inner horizon by a singular surface, when compared with the Reissner-Nordstr{\"o}m black hole\cite{Garfinkle:1990qj,Boulware:1986dr,Gibbons:1987ps}.  The solution also enjoys an $SL(2,\mathcal{R})$ symmetry, inherited from the S-duality in string theory, such that static dyonic solutions can be constructed \cite{Shapere:1991ta,Cheng:1993wp} as well as time-dependent ones \cite{Aniceto:2017gtx}.  A peculiar feature of this dynamic dilaton black hole is that it can be destroyed by spherically-symmetric and electrically-charged null fluid, therefore the censorship could be violated.  However, we remark that this violation is under the condition that all weak, dominant or strong energy conditions for the dilaton stress-energy tensor alone are violated sufficiently close to the apparent horizon.  One might suspect that this nontrivial dilaton configuration were responsible for such violation and wonder whether it would still happen for {\sl ordinary} matter.  In this article, we shall test the censorship in the old-fashioned way {\sl a la} Wald.  In contrast to Wald's conclusion that cosmic censorship is respected for extremal black holes in a canonical background without dilaton field\cite{Wald:1974wa,Wald:1997wa}, we find that a static electrically-charged extremal black hole in the GMGHS solution can be destroyed by a test charged particle if backreaction or self-energy(force) is ignored.  To our interest, after we introduce backreaction or self-energy(force) as suggested in the Hoop Conjecture, censorship is safely restored and, as a bonus, we find that the Weak Gravity Conjecture is necessary to satisfy the second law of black hole thermodynamics.  

This paper is organized as follows: the GMGHS solution of static black hole is reviewed in Sec.\ref{sec:static black hole}.  In Sec.\ref{sec:destroy black hole}, we show the extremal black hole could be destroyed by a test particle of specific energy, if backreaction is ignored.  However, we prove that censorship is recovered by using the Hoop Conjecture in Sec.\ref{sec:hoop conjecture}.  Moreover, in Sec.\ref{sec:weak gravity} we find that Weak Gravity Conjecture is necessary for the merging process.  We extend our discussion to the rotating case in Sec.\ref{sec:rotating black hole}.  At last, we have discussion in Sec.\ref{sec:conclusion}.

\section{Static black holes}\label{sec:static black hole}

We consider a static black hole solution in the Einstein-Maxwell-dilaton theory \cite{Aniceto:2017gtx}

\begin{eqnarray}\label{eqn:solution}
&&ds^2 = -(1-\frac{2M}{r}) dt^2 + (1-\frac{2M}{r})^{-1} dr^2 + r^2 (1-\frac{2D}{r}) (d\theta^2+\sin^2 \theta d\phi^2),\\
&&F=-\frac{Q}{r^2} dt \wedge dr,\\
&&e^{2\phi} = e^{2\phi_0} (1-\frac{2D}{r})
\end{eqnarray}

with the constraint

\begin{equation}
e^{-2\phi_0}Q^2 = 2MD
\end{equation}

This solution describes a massless scalar field coupled with the abelian vector field for a specific choice of scalar-Maxwell coupling $a=1$ in the action (\ref{eqn:action}).  $\phi_0$ is the value of dilaton field at the asymptotics, and we will set $\phi_0=0$ in the following discussion.  The black hole solution exists as long as censorship condition $M>D$ is satisfied.  We shall refer the extremal limit to the case $Q^2=2M^2$, where the singular radius coincides with the horizon.  The solution also exhibits an $SL(2,R)$ symmetry at the classical level, which is inherited from the S-duality in string theory.  With that being said, one may introduce a complex-valued scalar $\lambda\equiv \chi + e^{-2\phi}$ and (anti-) self-dual Maxwell field $F_{\pm}\equiv F \pm i^{\star}F$, where $\chi$ is known as the axion field.   Then the Lagrangian (\ref{eqn:action}) and solution (\ref{eqn:solution}) can be promoted to generate the dyonic solution with both electric and magnetic charges \cite{Shapere:1991ta,Cheng:1993wp}.

One may prefer the horizon to be a conventional two-sphere by adopting the areal coordinate, $R^2\equiv r(r-2D)$.  The singular surface at $r=2D$ corresponds to the origin $R=0$ in this new coordinate.  Away from the extremal limit, say $R>M\gg D$, the metric $g_{tt}$ can be expanded as 

\begin{equation}
-g_{tt}\simeq 1-\frac{2M}{R}+\frac{e^{-2\phi_0}Q^2}{R^2}+{\cal O}(R^{-3})
\end{equation}

and this black hole appears like a Reissner-Norstr{\"o}m black hole at far distance.
\section{Destroying a black hole}\label{sec:destroy black hole}

We consider the black hole is probed by a test particle with mass $m$ and charge $q$.  We only concern the probe limit for the moment, namely $m\ll M$ and $q\ll Q$.  The energy for a test particle along geodesic reads 

\begin{equation}
E = -q A_t + \sqrt{|g_{tt}|\bigg(m^2+\frac{\l^2}{g_{\phi\phi}}\bigg)},
\end{equation} 

for given angular momentum $\l$. A test particle needs at least energy $E_{L}$ to arrive at horizon,

\begin{equation}
E\ge E_{L} = \frac{qQ}{2M}.
\end{equation}

On the other hand, to destroy the black hole, we further assume that singularity becomes naked after the test particle is taken in, that is,

\begin{equation}
M+E \le \frac{(Q+q)^2}{2(M+E)}.
\end{equation}

This gives us a upper bound $E_{U}$.  After small $q$ expansion, one obtain

\begin{equation}
E\le E_{U}=E_{L}+\frac{1}{4M}(1-\frac{Q^2}{2M^2})q^2+{\cal O}(q^3)
\end{equation}

\begin{figure}[tbp]
	\includegraphics[width=0.6\textwidth]{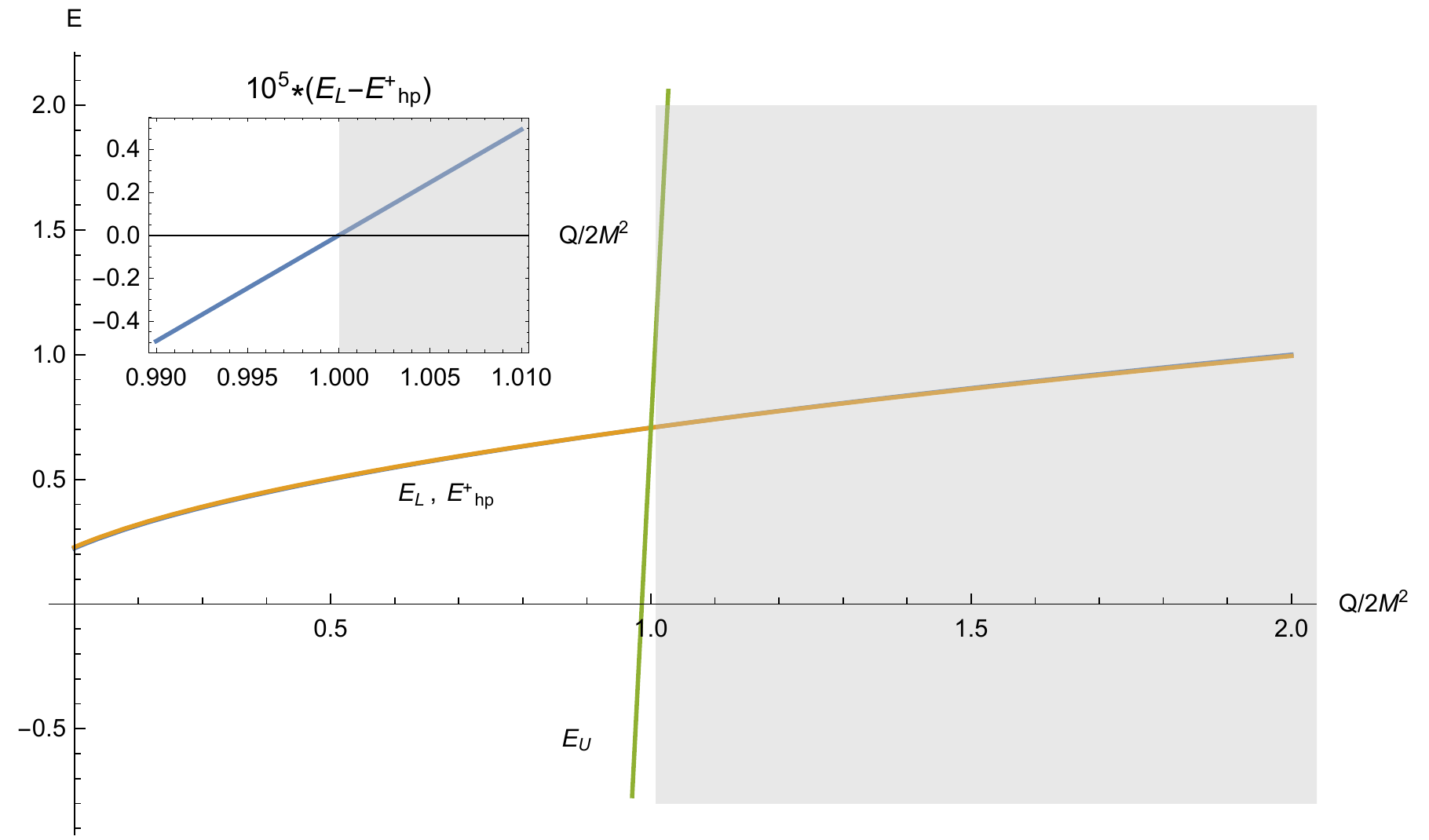}
	\caption{\label{fig:energy} Numeric shows that $E_U <  E_L < E^+_{hp}$ in the non-extremal black hole ($Q^2/2M^2<1$) and coincides only in the extremal limit.  The shaded regions are forbidden as censorship is violated.}
\end{figure}

The censorship condition before merging assures a positive coefficient for $q^2$ term.  There appears a possible energy window $E_{L} \le E \le E_{U}$ for the test particle to arrive at horizon and destroy the black hole.  However, it can be shown numerically that the actual upper bound is much lower.  In fact, one observes the opposite; as shown in Figure \ref{fig:energy}, $E_{U}<E_{L}$ for a generic non-extremal case, i.e. $Q^2<2M^2$.  It is {\sl barely} possible to destroy the extremal black hole with a specific energy $E=E_{L}=E_{U}$.  

If one assumes the test particle comes all the way from infinity, one shall also check the effective potential $V_{eff}$ remains negative along the way, that is,

\begin{equation}
V_{eff}=-\dot{r}=g_{rr}^{-1}\bigg(1+\frac{\l^2}{m^2g_{\phi\phi}}+\frac{(E+qA_t)^2}{m^2g_{tt}}\bigg)
\end{equation}

Since we only consider the static black hole solution, the test particle carries no angular momentum.  This gives us a negative definite effective potential as long as $E>m\sqrt{-g_{tt}}$.  

In conclusion, a test particle of a specific energy $E=\frac{Qq}{2M}$ and vanishing angular momentum $\l=0$ could fall straight to an extremal black hole and possibly destroy it, exposing the naked singularity and censorship being violated.  We note that above discussion does not take into account the higher order $q$-terms, or self energy of a charged particle, backreaction to the black hole background, radiation due to acceleration and etc.  In the following, we reconsider the merging process while the black hole receives the backreaction through the Hoop Conjecture.

\section{Cosmic Censorship and Hoop Conjecture}\label{sec:hoop conjecture}

The Hoop Conjecture states that the horizon forms when a mass $M$ gets compacted into a region whose circumferences in every direction is bounded by $C=4\pi M$.  According to the conjecture, a test particle with energy $E$ has become part of the black hole at the hoop radius $r_{hp}=2(M+E)$.  As a result, the lower bound $E^\prime_L$ and its small $q$ expansion become

\begin{equation}
E^{\prime}_L=\frac{(Q+q)q}{r_{hp}} \simeq \frac{Qq}{2M} + \frac{1}{2M}\bigg(1-\frac{Q^2}{2M^2}\bigg)q^2+{\cal O}(q^3)
\end{equation}

It is obvious that $E^{\prime}_L > E_U$ so there is, in fact, no energy window to destroy the black hole.  On one hand, the Hoop Conjecture can be regarded as the backreaction since it allows test particle energy to contribute part of the background.  On the other hand, the $q^2$ term in $E^{\prime}_L$ can also be interpreted as the contribution from self-force \cite{Smith:1980tv}.  Therefore, we may restate our conclusion that the static black hole in Einstein-Maxwell-dilaton theory can not be destroyed by a test particle with vanishing angular momentum if backreaction or self-force of the charged particle is considered.  The singularity remained hidden behind the horizon and censorship is protected.

\section{Connection to Weak Gravity Conjecture}\label{sec:weak gravity}

\begin{figure}[tbp]
	\includegraphics[width=0.7\textwidth]{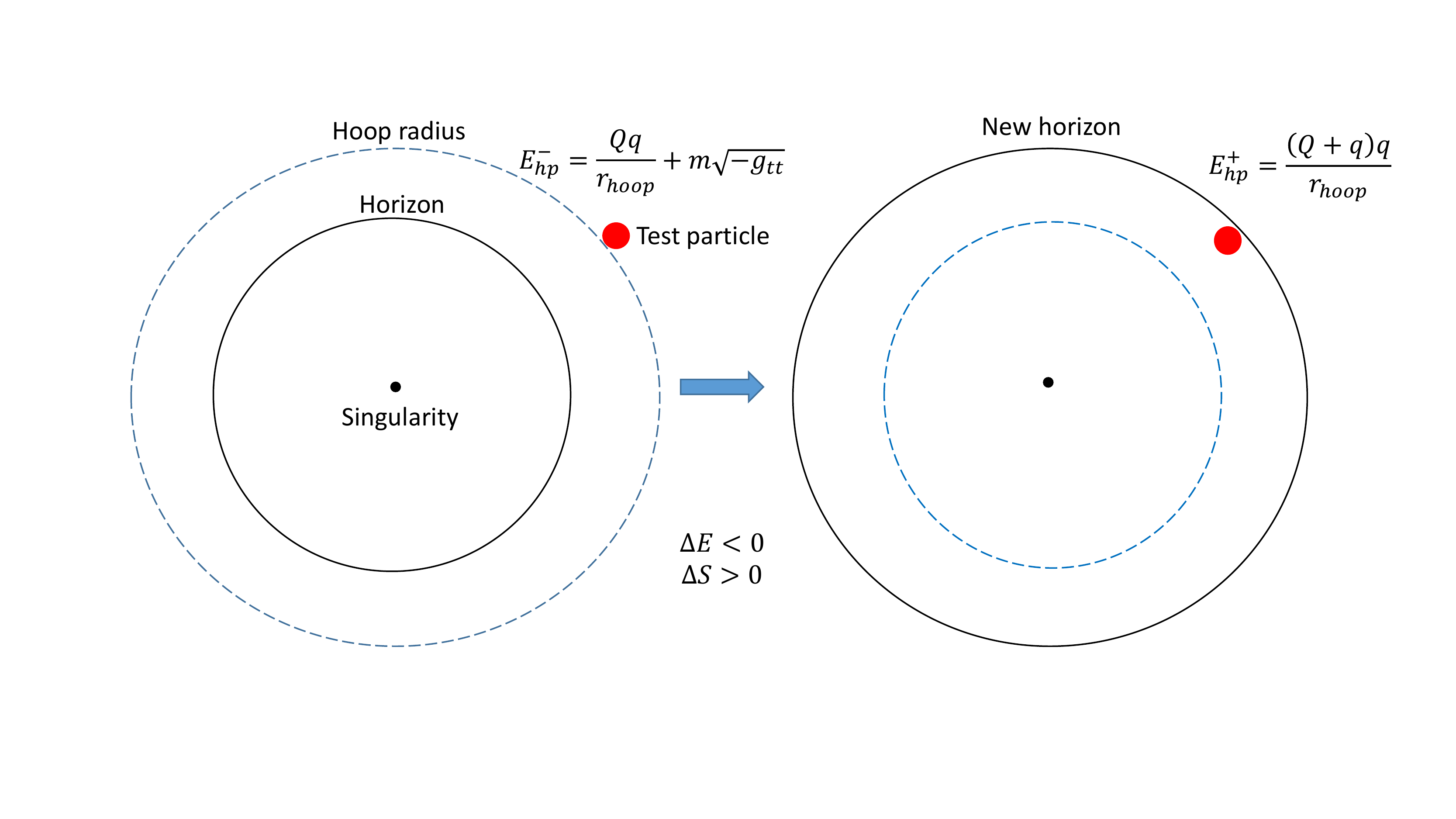}
	\caption{\label{fig:merge} We refine our merging process in two steps: before (left figure) and after (right figure) a test particle enters the reach of hoop radius.  We will show that energy is loss during the process, but the process is only energetic favored if the Weak Gravity Conjecture is imposed.}
\end{figure}

The computation in previous section is done right after the test particle steps into the reach of hoop radius.  As sketched in the Figure \ref{fig:merge}, one can refine the calculation by study the abrupt change of physical quantities across the hoop radius at some time $t_{hp}$.  If the test particle has a size of $L$, it would take some finite time $\epsilon = L/\dot{r}$ to completely cross the hoop radius.  In the following discussion, we will refer $E^{+}_{hp}$ as the energy of test particle at time $t_{hp}^{+}=t_{hp}+\epsilon/2$, which has been calculated as $E^{\prime}_L$.  However, right before it merges into the black hole, say at $t_{hp}^{-}=t_{hp}-\epsilon/2$, energy of the test particle is given by

\begin{equation}
E^{-}_{hp}=\frac{Qq}{r_{hp}}+m\sqrt{1-\frac{2M}{r_{hp}}}.
\end{equation} 

It is straightforward to show that $E^{-}_{hp} > E^{+}_{hp}$.  The energy order $E_U<E_{hp}^{+}=E^{\prime}_L<E_{hp}^{-}$ simply reassures that the black hole cannot be destroyed by a test particle.  The energy loss $|E^{-}_{hp} - E^{+}_{hp}|$ might take form of electromagnetic or gravitational radiation as the test particle flies across the hoop radius.  Despite the energy loss, one may wonder whether this process can happen naturally.  To answer this question, we shall compute the entropy change. The entropy change of black hole due to merging is given by 

\begin{equation}
\Delta S_{BH} = \pi r_{hp}^2 - \pi (2M)^2 = 8\pi M E^{+}_{hp} + 4\pi (E^+_{hp})^2
\end{equation}

\begin{figure}[tbp]
	\includegraphics[width=0.6\textwidth]{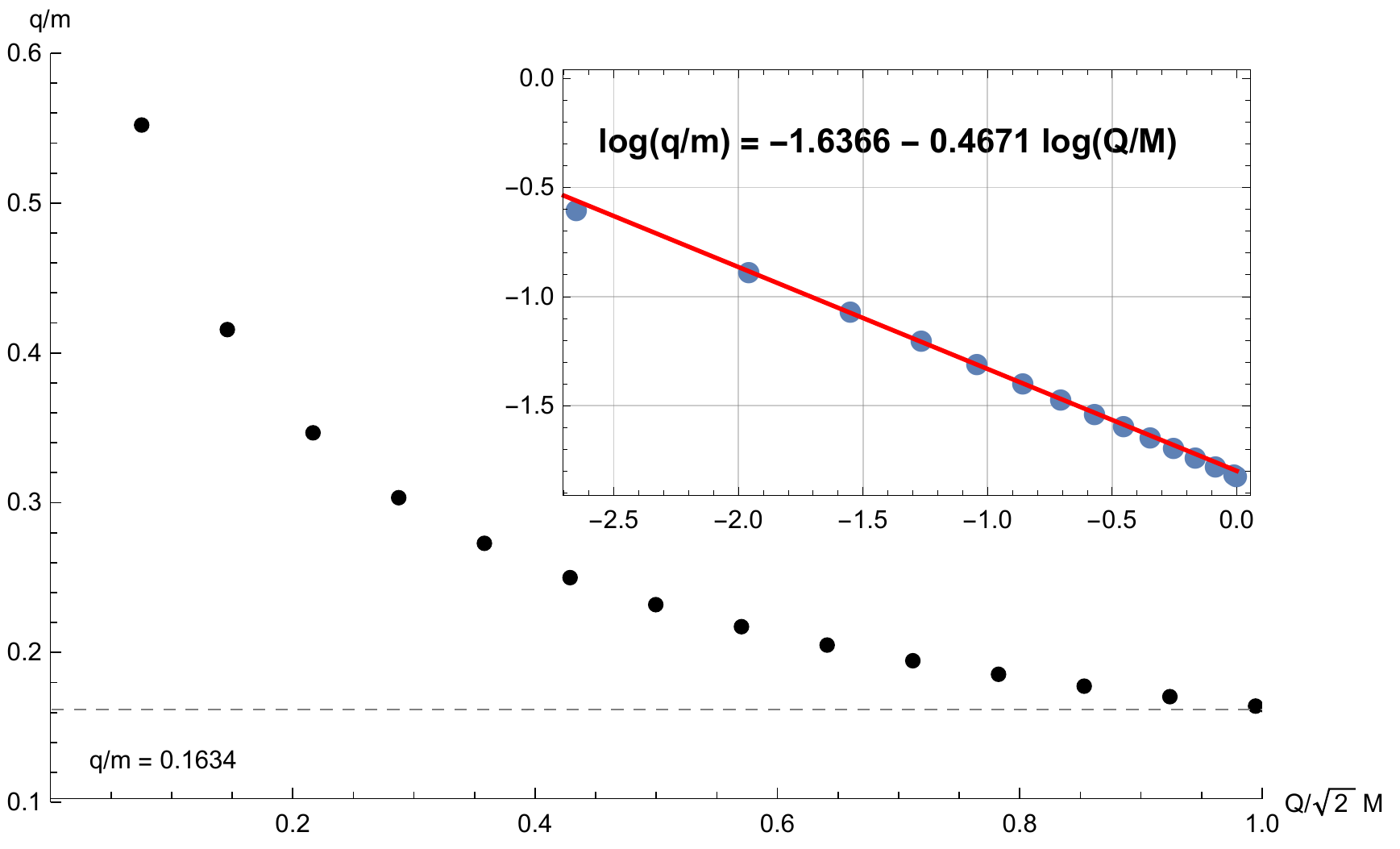}
	\caption{\label{fig:qmr} This plot shows that the minimum value $k$ of $q/m$ for a test particle changes against the black hole parameter $Q/\sqrt{2}M$.  $k$ varies from $1$ for the neutral black hole, to $0.1634$ for the extremal black hole.  Inset graph shows their log relation and numeric fit $k \propto (\frac{Q}{M})^{-0.4671}$.}
\end{figure}

In addition, the entropy change due to the test particle across the hoop radius is given by 

\begin{equation}\label{eqn:particle_entropy}
\Delta S_{p} = \int_{r_{hp}^-}^{r_{hp}^+} \frac{dQ}{T} \simeq \frac{E^{+}_{hp}}{T^{+}} - \frac{E^{-}_{hp}}{T^{-}} = 8\pi(M+E^+_{hp})E^+_{hp} - 8\pi M E^-_{hp},
\end{equation}
where $T^{+}=(4\pi r_{hp})^{-1}$ and $T^{-}=(8\pi M)^{-1}$ are the Hawking temperatures at time $t_{hp}^{\pm}$.  Here we have assumed that the energy loss of test particle is only carried away by thermal radiation, and $|t_{hp}^{+}-t_{hp}^{-}|=\epsilon \gg 8\pi M$, such that the black hole and test particle have enough time to reach thermal equilibrium during the process of merging.  As a result, the total entropy increase during merging process reads

\begin{equation}
\Delta S_{merge} = \Delta S_{BH}+ \Delta S_{p} = 4\pi (4M+3E^{+}_{hp})E^+_{hp} - 8\pi M E^{-}_{hp}.
\end{equation}

The generalized second law (GSL) of black holes thermodynamics suggests that $\Delta S_{merge} > 0$ for an irreducible process\cite{Bekenstein:1974ax}.  It is interesting to learn that to have $\Delta S_{merge} \ge 0$, one arrives at a lower bound for the mass-charge ratio of test particle, that is $q/m \ge k$ (in the unit $m_p=1$) for some constant $k$.  In Figure \ref{fig:qmr}, we found the numeric fit $k\propto (\frac{Q}{M})^{-0.4671}$ for non-extremal black hole. This surprising result agrees with the Weak Gravity Conjecture\cite{ArkaniHamed:2006dz}, which states, in its simplest form, that in a $U(1)$ gauge field coupled consistently to gravity, there must exist at least a particle whose proper mass is bounded by its charge.   Since the GSL should apply to all types of massive charged particles, we arrive at a stronger version of Weak Gravity Conjecture that the lower bounds for mass-charged ratio exist for all massive charged particles.\footnote{The authors thank the referee to point out this subtle difference in our version of Weak Gravity Conjecture.}  We remark that a recent proof of the Weak Gravity Conjecture by Hod \cite{Hod:2017uqc} using the universal relaxation bound\cite{Hod:2006jw}, which is also closely related to the GSL.

\section{Slowly rotating black hole}\label{sec:rotating black hole}

Now we would like to generalized our result to a test particle with nonzero angular momentum.  Accordingly, we have to modify the static black hole solution to a rotating one.  A slowly rotating black hole solution is given by \cite{Horne:1992zy}

\begin{eqnarray}
&&ds^2=-(1-\frac{r_+}{r})dt^2+(1-\frac{r_+}{r})^{-1}dr^2 + r^2(1-\frac{r_-}{r})d\Omega -2af(r)\sin^2{\theta}dtd\phi, \nonumber\\
&&f(r)=\frac{2r^2}{r_{-}^2}(1-\frac{r_{-}}{r})\log(1-\frac{r_-}{r})-1+\frac{2r}{r_{-}}+\frac{r_{+}}{r},\nonumber\\
&&e^{2\phi}=1-\frac{r_-}{r}, \nonumber\\
&&A_t=-\frac{Q}{r}, \ A_\phi = a\sin^2{\theta} \frac{Q}{r^2}
\end{eqnarray}

For slow rotation, we only keep terms up to order ${\cal O}(a)$.  This gives a magnetic potential $A_{\phi}$ and metric component $g_{t\phi}$ but $r_{+}=2M$ and $r_{-}=2D$ stays intact.  The angular momentum of black hole $J$ can be read off from $g_{t\phi}$ as $r\to \infty$, that is

\begin{figure}[tbp]
	\includegraphics[width=0.6\textwidth]{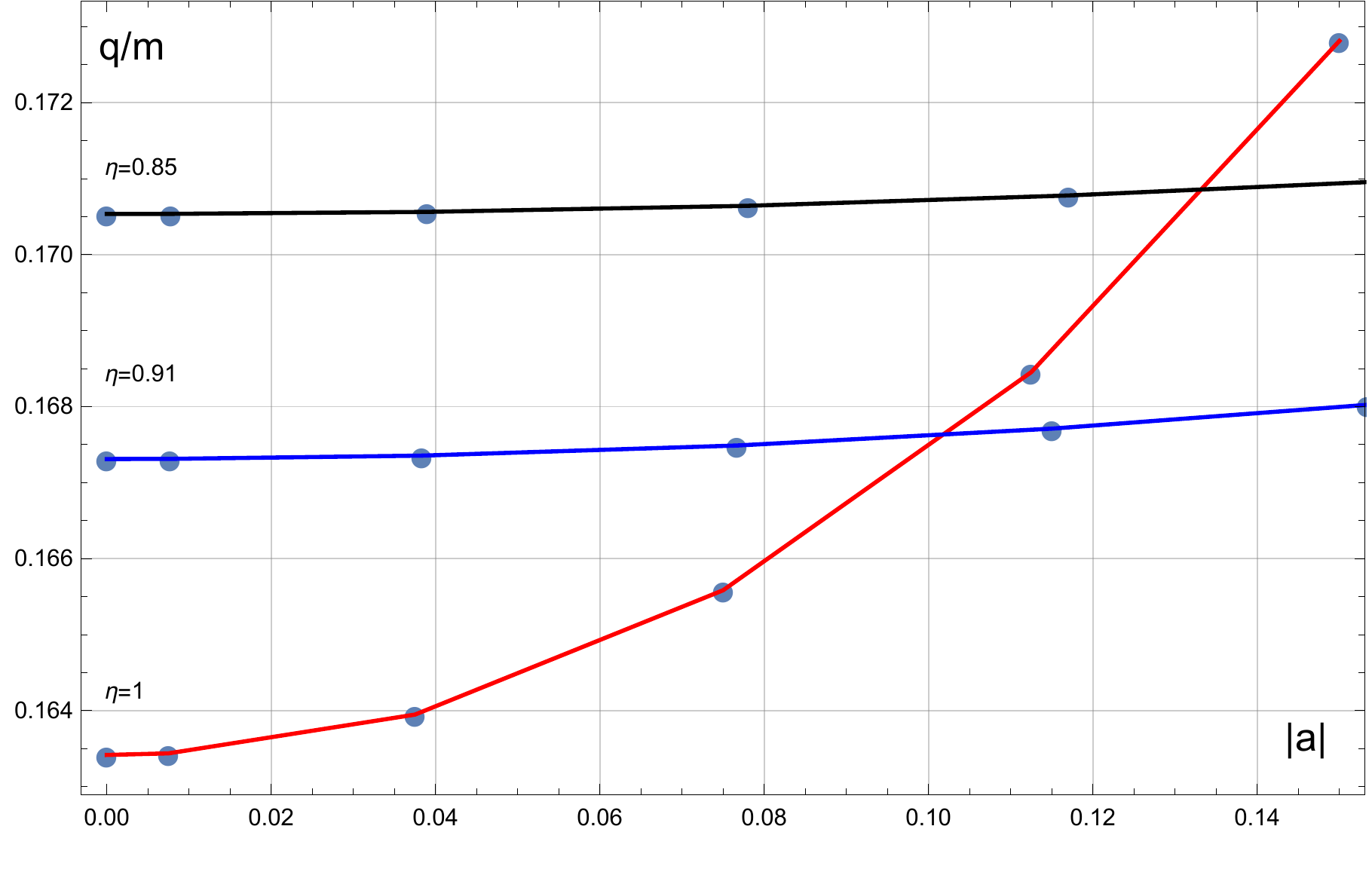}
	\caption{\label{fig:rotation} This plot shows that the minimum value of $q/m$ for a test particle changes against the black hole rotation parameter $a=\l/(M+\frac{D}{3})$.  In general, $k$ gradually increases with $|a|$.  The increase rate reaches the highest in the extremal limit ($\eta \equiv \frac{Q^2}{2M^2} \to 1$).}
\end{figure}

\begin{equation}
J = a(M+\frac{1}{3}D).
\end{equation}

We now reconsider the merging process but for a test particle with nonvanishing angular momentum $\l$.  We still begin with a static black hole and it starts to spin after intaking the test particle. The slowly rotating solution is justified by small parameter $a\sim \l/M \ll 1$.  Following the previous two-step approximation, we recalculate its energy before and after entering the range of hoop radius as follows:

\begin{eqnarray}
&&E_{hp}^{-}=\frac{Qq}{r_{hoop}}+\sqrt{(1-\frac{2M}{r_{hoop}})(m^2+\frac{\l^2}{g_{\phi\phi}})},\nonumber\\
&&E_{hp}^{+}=\frac{(Q+q)q}{r_{hoop}}+\frac{g_{t\phi}}{g_{\phi\phi}}\big(qA_{\phi}-\l\big)\big|_{r_{hoop}}
\end{eqnarray} 

To reach the hoop horizon, the minimal energy required for a test particle with nonzero angular momentum $\l$ is always higher than that with $\l=0$; on the other hand, the upper bound $E_U$ remains the same for small $a$.  Therefore, the black hole still cannot be destroyed by a test particle with nontrivial angular momentum.  In Figure \ref{fig:rotation}, it shows that the lower bound $k$ of charge-mass-ratio gradually increases with $|a|$.  The increase rate is highest in the extremal case and decreases as the black hole is away from extremality.

\section{Conclusion}\label{sec:conclusion}

In this paper, we study the censorship in the Einstein-Maxwell-dilaton theory by performing Wald's thought experiment.   We find naively that a static charged dilaton black hole could be destroyed by a test particle with a specific energy.  Fortunately, this can never happen if one considers the Hoop Conjecture, which effectively includes the backreaction in the background metric or in other words, self-force of the test particle in the vicinity of black hole horizon.    We further refine our calculation by break the merging process in two steps.  We find that in order for the process to be energetically favored, say, to have  non-decreasing entropy, the charge to mass ratio for the test particle has to be bounded from below, in the same spirit as the Weak Gravity Conjecture.  The charge-mass-ratio bound persists in rotating black holes and increases with rotation parameter $|a|$.  

We have several remarks.  First, the Weak Gravity Conjecture was discussed in many contexts, for instance, the discharging process of black holes \cite{Banks:2006mm}.  For completion, we show the charge-mass-ratio bound during discharging process in the appendix.  Nevertheless, we are the first to make its connection to the Hoop Conjecture in the merging process\footnote{A different version of Hoop Conjecture has proven useful to resolve the apparent violation of the generalized second law in the near-extremal Reissner-Nordstr{\"o}m (RN) black holes\cite{Hod:2015iig}.  Though its connection to the Weak Gravity Conjecture was not mentioned there.  Furthermore, the modified conjecture was specifically designed for RN black holes therefore inapplicable here.}.  A recent numeric study showed a close connection between the Weak Gravity Conjecture and censorship in the Einstein-Maxwell-($\Lambda$) theory with asymptotically anti-de Sitter boundary conditions\cite{Horowitz:2016ezu,Crisford:2017zpi,Crisford:2017gsb}.  Our discovery may contribute to a remarkable {\sl triangular} connection among the Weak Gravity Conjecture, the Weak Cosmic Censorship Conjecture and the Hoop Conjecture.  Second, the condition for thermal equilibrium during the merging process can be restated as $L \gg \alpha M |V_{eff}(r_{hoop})|$, where we have recovered the metric unit with $\alpha=\sqrt{\frac{G^2}{\hbar c^9}}\sim 10^{-34} m/J^2$.  Although this condition seems easy to be satisfied thanks to small value of $\alpha$, we argue that the result from equation (\ref{eqn:particle_entropy}) may still be valid even the condition fails. This is because entropy is a state function and temperature is well defined before and after the merging process.  Third, we observe that the censorship is protected in static or stationary black holes, however, we are not ready to argue against the censorship violation in the \cite{Aniceto:2017gtx} unless our discussion can be generalized to the dynamic dilaton solutions.  Fourth, our discussion on rotating black holes is restricted to small parameter $a$.  It would be nice if an exact Kerr black hole solution can be constructed and tested for the censorship in the Einstein-Maxwell-dilaton theory.  At last, we recall that some recent studies suggested the absence of horizon upon formation of dynamic (time-dependent) black holes \cite{Kawai:2013mda,Ho:2015vga}.  It is curious to see whether censorship violation in the \cite{Aniceto:2017gtx} can be simply resolved in the absence of horizon.  We leave the above-mentioned issues for future study.

\begin{acknowledgments}
	Part of this work has been reported in the 3rd LeCosPA symposium and physics department seminar at the National Taiwan University.  We are grateful to the useful comments by Chi-Te Liang and Gary Shiu.  This work is supported in part by the Taiwan's Ministry of Science and Technology (grant No. 106-2112-M-033-007-MY3).
\end{acknowledgments}

\appendix

\section{Weak Gravity Conjecture and black holes discharge}	
In this appendix, we will derive the lower bound of charge-to-mass ratio suggested by the discharging process.  Following the discussion in \cite{Banks:2006mm}, we will assume the extremal dilaton black hole can still emit charged particle spontaneously\footnote{However, one may also regard the discharging process less efficiently as simply quantum tunneling\cite{Parikh:1999mf}.  This suggests no obvious role played by the Weak Gravity Conjecture, as pointed out in \cite{Furuuchi:2017upe}.}.  There are two cases in consideration.  At high temperature or small black holes regime, one considers emission as thermal radiation with a chemical potential.  The emission rate reads

\begin{equation}
e^{-\frac{1}{T}(m-\frac{qQ}{r_+})}
\end{equation}

which is dominant if the exponent becomes positive.  This gives a lower bound $q/m \ge \sqrt{2}$.  On the other hand, at low temperature or large black holes regime, one considers the Schwinger process.  The emission rate becomes

\begin{equation}
e^{-\frac{\pi m^2 r_+^2}{qQ}}
\end{equation}

It starts to discharge appreciably when the exponent is less than $1$.  This gives another bound $q/m^2 \ge 4\pi M/ \sqrt{2}$.

\end{document}